\documentclass[aps,prd,groupedaddress,showpacs]{revtex4-1}

\usepackage{amssymb}
\usepackage{amsmath}
\usepackage{bbm}
\usepackage{color}
\usepackage{graphicx}

\DeclareMathOperator{\diracd}{\delta}
\DeclareMathOperator{\im}{i}

\DeclareMathOperator{\f}{f}

\DeclareMathOperator{\eqdef}{\stackrel{\scriptscriptstyle\wedge}{=}}

\def\gtwid{{\,\raise.3ex\hbox{$>$\kern-.75em\lower1ex\hbox{$\sim$}}\,}}
\def\ltwid{{\,\raise.3ex\hbox{$<$\kern-.75em\lower1ex\hbox{$\sim$}}\,}}

\begin{document}

\title{Colorful SU(2) center vortices in the continuum and on the lattice}

\author{Thomas Schweigler}
\author{Roman H\"ollwieser\footnote{Funded by Austrian Science Fund (FWF) under contract P22270-N16.}}
\email{hroman@kph.tuwien.ac.at}
\author{Manfried Faber}
\email[]{faber@kph.tuwien.ac.at}
\affiliation{Institute of Atomic and Subatomic Physics, Vienna University of Technology, Wiedner Hauptstr. 8-10, 1040 Vienna, Austria}
\author{Urs M. Heller}
\affiliation{American Physical Society, One Research Road, Ridge, New York 11961, USA}

\date{\today}

\begin{abstract}
The spherical vortex as introduced in [Phys.\ Rev.\ {\bf D77}, 014515 (2008)]
is generalized. A continuum form of the spherical vortex is derived and
investigated in detail. The discrepancy between the gluonic lattice
topological charge and the index of the lattice Dirac operator described in
previous papers is identified as a discretization effect. The importance
of the investigations for Monte Carlo configurations is discussed.
\end{abstract}


\maketitle

\section{Introduction}

The distribution of topological charge density is closely linked to the
phenomena of the axial anomaly and spontaneous chiral symmetry breaking.
There are strong hints that the QCD vacuum is dominated by center vortices
\cite{'tHooft:1977hy,Vinciarelli:1978kp,Yoneya:1978dt,Cornwall:1979hz,Mack:1978rq,Nielsen:1979xu}.
Center vortices can explain color confinement
\cite{DelDebbio:1996mh,DelDebbio:1997ke,Kovacs:1998xm,Greensite:2003bk} and
seem to be also of paramount importance for spontaneous chiral symmetry
breaking
\cite{Reinhardt:2000ck,deForcrand:1999ms,Alexandrou:1999vx,Engelhardt:2002qs,Hollwieser:2008tq}.
Center vortices can contribute to the topological charge density through
intersection and writhing points \cite{Reinhardt:2000ck}, but also through
their color structure. The prototype of the later contribution in SU(2)
gauge theory is the spherical vortex introduced in previous articles of our
group \cite{Jordan:2007ff,Hollwieser:2010mj,Hollwieser:2012kb}. In this
paper, a generalization of the original spherical vortex is constructed.
Subsequently, the generalized spherical vortex is investigated in
detail on the lattice and in the continuum.

In section \ref{sec:original}, we start with a description of the original spherical
vortex on the lattice as introduced in \cite{Jordan:2007ff}. From this
lattice gauge configuration, we then derive the corresponding continuum
object. With the continuum form, a more general spherical vortex can be
described. The action and topological charge density of such generalized
spherical vortices in the continuum are discussed in section
\ref{sec:action} and \ref{sec:topdencont}. We show that the large
action of the original spherical vortex encountered in previous
investigations can be significantly reduced by spreading the vortex over
several time slices.
In section \ref{sec:gen_latt}, a generalized spherical vortex on the
lattice is derived from the continuum form. Subsequently its gluonic and
fermionic properties are investigated. In particular, we show that the
discrepancy between the gluonic lattice topological charge and the index of
the lattice Dirac operator seen for the original spherical vortex is simply
a discretization effect. We conclude with remarks on the contributions of
colorful center vortices to topological charge and chiral symmetry breaking
in Monte Carlo configurations.



\section{The original spherical vortex}   
\label{sec:original}

The original spherical vortex on the lattice as introduced in
\cite{Jordan:2007ff} is given by the lattice links \footnote{Note that
throughout this paper Einstein's summation convention is used. As already
mentioned, we will work with SU(2) gauge theory. The Pauli matrices will be
denoted by $\sigma$. Throughout this paper, the lattice constant is set to
$a=1$.}
\begin{align}
\begin{split}
	U_{i} (x) &= \mathbbm{1}  \\
	U_{4} (x) &=
\begin{cases}
 \cos  \left[ \alpha\left( \left| \vec{r} - \vec{r}_0 \right| \right) \right]
  \mathbbm{1}  - i \ \vec{e_r} \cdot \vec{\sigma} \
  \sin \left[ \alpha\left( \left| \vec{r} - \vec{r}_0 \right| \right) \right]
& \mbox{for} \quad t = 1 \\
 \mathbbm{1} & \mbox{else}
\end{cases}.
\label{originallinks}
\end{split} 
\end{align}
%
Here $x=\{ x_1, x_2, x_3, t\}$ stands for the lattice site and $\vec{r}$
denotes the spatial components only. The unit vector pointing from the
vortex midpoint $\vec{r}_0$ to the spatial lattice site $\vec{r}$ is
denoted by
\begin{equation*}
\vec{e_r} = \frac{\vec{r} - \vec{r}_0}{\left| \vec{r} - \vec{r}_0 \right|}  
\label{def_rn}.
\end{equation*}
$U_{\mu}(x)$ is the link connecting the lattice site $x$ with the site
$x + \hat{\mu}$, where $\hat{\mu}$ is the unit vector pointing in the $\mu$
direction. The gauge links on the lattice represent the adjoint gauge
transporters and, therefore, transform like $U_{\mu} (x) \to
\Omega(x + \hat{\mu}) \ U_{\mu} (x) \ \Omega(x)^{\dagger}$ under a gauge
transformation $\Omega(x)$. Indices denoted by Latin letters run from 1 to
3 (spatial indices), while indices denoted by Greek letters run from 1 to 4
(spatial plus temporal indices). The function $\alpha(r)$ rises linearly
from $-\pi$ to 0, {\it i.e.},
\begin{equation}
	\alpha(r) = -\pi \cdot
\begin{cases}
 1 & \mbox{for} \ r < R - \frac{d}{2} \\
 \frac{R + \frac{d}{2} - r}{d} & \mbox{for} \
   R - \frac{d}{2} \leq r \leq R + \frac{d}{2} \\
 0 & \mbox{for} \ r > R + \frac{d}{2}
\end{cases}.
\label{alpha}
\end{equation}

Note that the spherical vortex configuration posses only non-trivial
temporal links $U_4$. All spatial links $U_i$ are trivial and therefore the
gluonic lattice topological charge $Q=0$. On the other hand the index of
the Dirac operator $\mbox{ind}\left[ D \right] = n_- - n_+$, defined as the
difference between the number of left- and right-handed chiral zeromodes,
is always 1. This gives a discrepancy between the index and the gluonic
topological charge. Note that no violation of the index theorem
$\mbox{ind}\left[ D \right] = Q$ occurs as the theorem is only stated for
the continuum limit. When cooling, the gluonic topological charge
approaches 1 and the discrepancy $\mbox{ind}\left[ D \right] \neq Q$ gets
resolved \cite{Jordan:2007ff,Hollwieser:2010mj,Hollwieser:2012kb}.

Naively, one would perform the continuum limit of the spherical vortex by
simply splitting the temporal links and leaving the spatial links trivial.
This would give gauge configurations of the form
\begin{align}
\begin{split}
	U_i(x) &= \mathbbm{1},  \\
 	U_4(x) &= \cos  \left[w(t) \ \alpha\left( \left| \vec{r} - \vec{r}_0 \right| \right) \right]  \mathbbm{1}
  - i \ \vec{e_r} \cdot \vec{\sigma} \ \sin \left[ w(t) \  \alpha\left( \left| \vec{r} - \vec{r}_0 \right| \right) \right]
 \quad \mbox{with} \quad \sum_{t}{ w(t)} = 1 \, ,
\label{pospher_links_spread_naive}
\end{split}
\end{align}
where $w(t)$ determines how the temporal links are split. As one can
easily see, a singularity occurs at the vortex midpoint when $0 \neq w(t)
\neq 1$. Moreover, the gluonic topological charge $Q$ clearly remains 0
independent of $w(t)$ which doesn't agree with the cooling history of
the original vertex, Eq.~(\ref{originallinks}). Therefore, this simple
procedure seems not to be suitable for performing the continuum limit of
the lattice spherical vortex.

In order to find a better continuum limit and to gain understanding of the
spherical vortex, one has to apply a gauge transformation to the link
configuration given in Eq.~(\ref{originallinks}). With the lattice gauge
transformation
\begin{equation*}
	\Omega(x) = \begin{cases}
 g(\vec{r}) &\mbox{for} \quad 1 < t \leq t_g \\
 \mathbbm{1} &\mbox{else} \end{cases} ,
\end{equation*}
where
\begin{equation}
g(\vec{r}) = \cos  \left[  \alpha\left( \left| \vec{r} - \vec{r}_0 \right| \right) \right]  \mathbbm{1}
  + i \ \vec{e_r} \cdot \vec{\sigma} \ \sin \left[  \alpha\left( \left| \vec{r} - \vec{r}_0 \right| \right) \right] \, ,
\label{hgtrans}
\end{equation}
the spherical vortex becomes
\begin{align}
\begin{split}
	U_{i} (x) &= \begin{cases}
  g \left( \vec{r} + \hat{i} \right) \ g\left(\vec{r}\right)^{\dagger}
 & \mbox{for} \quad  1 < t \leq t_g \\
  \mathbbm{1} & \mbox{else}
 \end{cases} , \\
	U_{4} (x) &= \begin{cases}
  g\left( \vec{r} \right)^{\dagger} & \mbox{for} \quad t = t_g \\
  \mathbbm{1} & \mbox{else}
 \end{cases} .
\label{vvlattoriginal}
\end{split} 
\end{align}
{}From this it becomes clear that the spherical vortex represents a
transition (in the temporal direction) between two pure gauge fields. The
transition occurs between $t=1$ and $t=2$. No transition occurs between
$t=t_g$ and $t=t_g+1$ since all plaquettes there are trivial. For $t \leq
1$ the pure gauge field is trivial, {\it i.e.}, the winding number $n_w=0$.
For $t > 1$ the pure gauge field is generated by the hedgehog gauge
transformation, Eq.~(\ref{hgtrans}), which has $n_w=1$
\cite{Diakonov:2002fq}. This means that in the continuum limit, the
spherical vortex (representing a transition between a vacuum with winding
number 0 and 1) has topological charge $Q=1$ \cite{Diakonov:2002fq}. It can
be regarded as a squeezed instanton with center vortex structure.

Assuming an infinitely big temporal extent of the lattice and taking $t_g
\to \infty$, the continuum field corresponding to Eq.~(\ref{vvlattoriginal})
can be written as
\begin{equation}
\mathcal{A}_{\mu} =  \f \left( t \right) \ i \left(\partial_\mu g \right) g^\dagger
\label{vvtransgeneral} \, ,
\end{equation}
where $g$ is given in Eq.~(\ref{hgtrans}) and $\f (t)$, determining the
transition in temporal direction t, changes from 0 to 1 within one lattice
unit. Clearly, one could also use a function $\f (t)$ that changes more
slowly. In this case, the corresponding lattice object cannot be
represented by configurations of the form of Eq.~(\ref{originallinks}) anymore.
We refer to the construction with a smoother function $f(t)$ as a
generalized spherical vortex. One construction of such a generalized
spherical vortex on the lattice will be investigated in section
\ref{sec:gen_latt}. Before doing so, we discuss the action and topological
charge density of the continuum gauge field, given in Eq.~(\ref{vvtransgeneral}) in the next two sections.

 
\section{Action of the spherical vortex in the continuum}
\label{sec:action}

In this section, we calculate the action for the generalized spherical
vortex in the continuum given in Eq.~(\ref{vvtransgeneral}). For simplicity, we
set the vortex midpoint $\vec{r}_0$ to $0$ in this and the next
section. Moreover, we use the notation $|\vec{r}|=r$.

To explicitly evaluate the action $S$ for the $\mathcal{A}_{\mu}$ given in
Eq.~(\ref{vvtransgeneral}), we have to evaluate $i \left(\partial_\mu g
\right) g^\dagger$ where $g$ is given in Eq.~(\ref{hgtrans}). To do so, we
first have a look at $i \left(\partial_\mu g \right) g^\dagger$ for a
general gauge transformation
\begin{equation*}
 g = q_0 \ \sigma_0 + \im \vec{q} \cdot \vec{\sigma} \, ,
\end{equation*}
where $q_0$ and $\vec{q}$ are constraint by $q_0^2 + \left| \vec{q} \right|^2
= 1$. Using the identity $\sigma^a \sigma^b = \diracd^{a b} \mathbbm{1} +
\im \epsilon^{a b c} \sigma^c$ and
\[
 q_0 \partial_{\mu} q_0 + q_k \partial_{\mu} q_k = \frac{1}{2} \partial_{\mu} \left( q_0 q_0 + q_k q_k \right) = \frac{1}{2} \partial_{\mu} \left( 1 \right) = 0 \, ,
 \]
one gets, after a few lines of calculation,
\begin{equation}
 i \left(\partial_\mu g \right) g^\dagger
 = \vec{\sigma} \cdot \left[ \left( \partial_{\mu} q_0 \right) \vec{q} -
   q_0  \left( \partial_{\mu} \vec{q} \right) +
   \vec{q} \times \left( \partial_{\mu} \vec{q} \right) \right] \, .
\label{genpqqd}
\end{equation}
For the $g$ given in Eq.~(\ref{hgtrans}) we have $q_0 = \cos (\alpha (r) )$
and $\vec{q} = \vec{e_r} \ \sin \left( \alpha(r) \right)$. Inserting this
into Eq.~(\ref{genpqqd}) and multiplying with $ \f \left( t \right)$ gives
($\mathcal{A_{\mu}}  = \frac{\sigma^a}{2} A_{\mu}^a$)
\begin{align}
\begin{split}
	A_{i}^a = \ &\f \left( t \right) \cdot 2 \cdot \Bigg( \frac{x_i x_a}{r^3} \sin  \left(  \alpha(r) \right)  \cos \left(  \alpha(r) \right) + \epsilon_{i a k} \ x_k \ \frac{1}{r^2} \ \sin^2  \left(  \alpha(r) \right) \\
	&-\delta_{i a} \frac{1}{r} \sin  \left(  \alpha(r) \right) \cos \left(  \alpha(r) \right) - \frac{x_i x_a}{r^2} \alpha'(r) \Bigg) \, , \\
	A_{4}^a = \ & 0 \, .
\label{a_log}
\end{split}
\end{align}
In the following, we use the notation
\begin{equation}
	A_{\mu}^a = \f \left( t \right) \ A_{\mu +}^a \, .
\label{a_short}
\end{equation}

The field strength tensor is given by 
\begin{equation*}
	F_{\mu \nu}^a = \partial_{\mu} A_{\nu}^a - \partial_{\nu} A_{\mu}^a
 - \epsilon^{a b c} A_{\mu}^b A_{\nu}^c \, .
\end{equation*}
The different elements of the field strength tensor can be simplified
considerably for our particular gauge field configuration. Let us first
have a look at the elements of the form $F_{4 i}^a$. Using the fact that
the gauge field has no temporal components, {\it i.e.}, $A_{4}^a = 0$, we
can simplify these elements to
\begin{equation}
	F_{4 i}^a = \partial_4 A_i^a \, . \label{elsimpl}
\end{equation}
The $F_{i j}^a$ can be simplified too. With the notation (\ref{a_short})
we get
\begin{align}
\begin{split}
	F_{i j}^a &= \f \left( t \right) \partial_{i} A_{j +}^a - \f \left( t \right) \partial_{j} A_{i +}^a - \f \left( t \right)^2 \epsilon^{a b c} A_{i +}^b A_{j +}^c \\ 
	&= \f \left( t \right) \left( \partial_{i} A_{j +}^a - \partial_{j} A_{i +}^a - \epsilon^{a b c} A_{i +}^b A_{j +}^c \right) + \left( \f \left( t \right) - \f \left( t \right)^2 \right) \ \epsilon^{a b c} A_{i +}^b A_{j +}^c \\
	&= \f \left( t \right) \left( 1 - \f \left( t \right) \right) \ \epsilon^{a b c} A_{i +}^b A_{j +}^c \, .
\label{magsimpl}
\end{split}
\end{align}

Using these results, we can now evaluate the action. Let us first evaluate
$\mbox{tr}_C \left[ \mathcal{F}_{4 i} \mathcal{F}_{4 i} \right]$. With
(\ref{elsimpl}) and (\ref{a_short}) we get
\begin{equation*}
	\mbox{tr}_C \left[ \mathcal{F}_{4 i} \mathcal{F}_{4 i} \right] = \frac{1}{2} \ F_{4 i}^a F_{4 i}^a =  \frac{1}{2} \left( \frac{d}{d t} \f (t) \right)^2 A_{i +}^a A_{i +}^a \, .
\end{equation*}
Inserting the explicit form of $A_{i +}^a$ one can evaluate  
\begin{equation*}
	 A_{i +}^a A_{i +}^a = 4 \left( \frac{2}{r^2} \sin\left(\alpha(r)\right)^2 + \alpha'(r)^2 \right) \, .
\end{equation*}
Similarly we get
\begin{equation*}
	\mbox{tr}_C \left[ \mathcal{F}_{i j} \mathcal{F}_{i j} \right]
 = \frac{1}{2} \ F_{i j}^a F_{i j}^a = \f \left( t \right)^2
  \left( 1 - \f \left( t \right) \right)^2 \ \epsilon^{a b c} A_{i +}^b A_{j +}^c \
   \epsilon^{a d e} A_{i +}^d A_{j +}^e \, ,
\end{equation*}
where Eq.~(\ref{magsimpl}) has been used in the last step. With the help of
a computer algebra program, one can evaluate
\begin{equation*}
 	\epsilon^{a b c} A_{i +}^b A_{j +}^c \ \epsilon^{a d e} A_{i +}^d A_{j +}^e
 = \frac{32}{r^4} \sin\left(\alpha(r)\right)^2 \left( \sin\left(\alpha(r)\right)^2
  + 2 r^2 \alpha'(r)^2 \right) \, .
\end{equation*}
Combining these identities and switching to polar coordinates gives 

\begin{align}
\begin{split}
	S &= \frac{1}{2 g^2} \int d^4x \ \mbox{tr}_C \left[ \mathcal{F}_{\mu \nu} \mathcal{F}_{\mu \nu} \right]
 = \frac{1}{2 g^2} \int d^4x \left( \mbox{tr}_C \left[ \mathcal{F}_{i j} \mathcal{F}_{i j} \right]
   + 2 \ \mbox{tr}_C \left[ \mathcal{F}_{4 i} \mathcal{F}_{4 i} \right] \right) \\
	&=  \frac{1}{2 g^2} \Bigg( \int{ dt \ \f \left( t \right)^2 \left( 1 - \f \left( t \right) \right)^2 }
  \int{ dr \ \frac{64 \pi}{r^2} \sin\left(\alpha(r)\right)^2 \left( \sin\left(\alpha(r)\right)^2 + 2 r^2 \alpha'(r)^2 \right)} \\
	& \quad + \int{ dt \left( \frac{d}{d t} \f (t) \right)^2 } \int{ dr  \ 16 \pi \left( 2 \sin\left(\alpha(r)\right)^2 + r^2 \alpha'(r)^2 \right)} \Bigg) \, .
\label{skgeneralf}
\end{split}
\end{align}

So far, the calculation has been done for a general $\f \left( t \right)$.
Now we will choose $\f \left( t \right)$ as the piecewise linear function
\begin{equation}
	\f_{\Delta t}(t) =  \begin{cases} 0 & \mbox{for} \ t < 1 \\ \frac{t-1}{\Delta t} & \mbox{for} \ 1 \leq t \leq  1 + \Delta t \\ 1 & \mbox{for} \ t >  1+ \Delta t \end{cases} \label{fk_ramp},
\end{equation}
where $\Delta t$ stands for the duration of the transition. Inserting
$\f_{\Delta t}(t)$ into Eq.~(\ref{skgeneralf}) gives
\begin{align}
\begin{split}
	S =  \frac{1}{2 g^2} \Bigg( &\Delta t \cdot \int{ dr \ \frac{32 \pi}{15 r^2} \sin\left(\alpha(r)\right)^2 \left( \sin\left(\alpha(r)\right)^2 + 2 r^2 \alpha'(r)^2 \right)} \\
	+ \ &\frac{1}{\Delta t} \cdot \int{ dr  \ 16 \pi \left( 2 \sin\left(\alpha(r)\right)^2 + r^2 \alpha'(r)^2 \right)} \Bigg) \label{sktempdone} \, . 
	\end{split}
\end{align}
%

One can see from Eq.~(\ref{sktempdone}), that as long as none of the
spatial integrations gives zero, the action diverges for both $\Delta t \to
\infty$ and $\Delta t \to 0$. Note that the first term in
Eq.~(\ref{sktempdone}) represents the magnetic and the second term the electric
contributions to the action. This means that for $\Delta t \to 0$, the
action is purely electric, and for $\Delta t \to \infty$, it is purely
magnetic. In between, there is a minimum where the electric and magnetic
contributions are equal.

Performing the spatial integration is a difficult task, which was done with
a computer algebra program. The result for a general ratio of $d/R$ is
lengthy and therefore not explicitly given here. However, let us state
some properties of the result. If we minimize the action for a given $R$
and $d$ with respect to $\Delta t$, we get an expression that depends only
on the ratio $d/R$. This expression falls monotonically with $d/R$ in the
allowed range $0 \leq  d/R \leq 2$. It diverges for $d/R \to 0$ and
approaches its minimum for $d/R \to 2$. The absolute minimum of the action
(minimized with respect to $\Delta t$, $R$ and $d$) is given by $S_{min} =
1.667 \ S_{Inst}$ ($R=d/2\approx0.305\Delta t$). Here we denoted the action of one instanton by $S_{Inst}
= 1/(2 g^2) \ 16 \pi^2$. This action serves as a lower bound for objects
with topological charge $|Q| = 1$.
Fixing the ratio $d/R$ gives the action as a function of $R$ and $\Delta
t$. For $d/R=1$ we get
\begin{equation}
	\frac{S(\Delta t)}{S_{Inst}} = \frac{0.4358}{R} \cdot \Delta t + \frac{3.722 \ R}{\Delta t} \, .
\label{eval_action_ramp_fk}  
\end{equation}

\section{Topological charge density of the spherical vortex in the continuum}
\label{sec:topdencont}

In this section, we calculate the topological charge density $q(x)$ of
field configurations of the form of Eq.~(\ref{vvtransgeneral}). For now,
the calculation is done for a general gauge transformation $g$ and a
general $\f \left( t \right)$ with the restriction, that they have to be
chosen in such a way, that the resulting fields are free of singularities.
Moreover, we will assume that $g$ is independent of the temporal coordinate
t. In the following, we again use the shorthand notation $\mathcal{A}_{i +}
=i \left(\partial_i g \right) g^\dagger$.

As is well known \cite{Diakonov:2002fq}, $q(x)$ can be calculated as the
full derivative
\begin{equation*}
	q(x) = \partial_{\mu} K_{\mu} (x) \, , \quad \mbox{with}
 \quad K_{\mu} = \frac{1}{16 \pi^2} \epsilon_{\mu  \alpha \beta \gamma}
  \left(  A_{\alpha}^a \partial_{\beta} A_{\gamma}^a -
   \frac{1}{3} \epsilon^{a b c} A_{\alpha}^a A_{\beta}^b A_{\gamma}^c \right) \, .
\end{equation*}
Let us now split $\partial_{\mu} K_{\mu}$ into a spatial ($\partial_{i}
K_{i}$) and a temporal part ($\partial_{4} K_{4}$). First, we treat
the spatial part. With the assumption that $g$ is independent of the
temporal coordinate, one can easily see that the gauge field
(\ref{vvtransgeneral}) has no temporal component, i.e. $\mathcal{A}_4 = 0$.
Therefore, $K_i$ evaluates to zero:
\begin{align*}
\begin{split}
	K_i &= \frac{1}{16 \pi^2} \epsilon_{i  \alpha \beta \gamma}
 \left( A_{\alpha}^a \partial_{\beta} A_{\gamma}^a -
   \frac{1}{3} \epsilon^{a b c} A_{\alpha}^a A_{\beta}^b A_{\gamma}^c \right)
= \frac{1}{16 \pi^2} \ \epsilon_{i  \alpha 4 \gamma} \  A_{\alpha}^a
  \partial_{4} A_{\gamma}^a \\
&= - \ \frac{1}{16 \pi^2} \ \epsilon_{i  j k} \ A_{j}^a \partial_{4} A_{k}^a
= - \ \frac{1}{16 \pi^2} \ \epsilon_{i  j k} \ A_{j +}^a A_{k +}^a
  \f(t) \f'(t) = 0 \, .
\end{split}
\end{align*}
This means, that we can identify the topological charge density with the
temporal derivative of $K_4$, {\it i.e.},
\begin{equation*}
	q (x) = \partial_4 K_4 \, .
\end{equation*}
{}From the definition of $K_{\mu}$ given above, we see that we can write
$K_4$ as
\begin{equation*}
	K_4 = - \ \frac{1}{16 \pi^2}  \left( \f(t)^2 \ \epsilon_{i j k} \
	A_{i +}^a \partial_j A_{k +}^a - \f(t)^3 \ \frac{1}{3} \
	\epsilon_{i j k} \ \epsilon^{a b c} \ A_{i +}^a A_{j +}^b A_{k +}^c
	\right) \, ,
\label{k4wft}
\end{equation*}
where we have used $\epsilon_{4 i j k} = - \ \epsilon_{i j k}$. Using the
fact that the field strength vanishes for pure gauge fields, {\it i.e.},
$F_{\mu \nu}^a = \partial_{\mu} A_{\nu +}^a - \partial_{\nu} A_{\mu +}^a -
\epsilon^{a b c} A_{\mu +}^b A_{\nu +}^c = 0 $, one can rewrite the term
$ \epsilon_{i j k} \ A_{i +}^a \partial_j A_{k +}^a $ as $ \frac{1}{2} \
\epsilon_{i j k} \ \epsilon^{a b c} \ A_{i +}^a A_{j +}^b A_{k +}^c $.
Using this result, we get
\begin{equation}
	K_4 = - \ \frac{1}{16 \pi^2}  \left( \left( \frac{1}{2} \f(t)^2 -
 \frac{1}{3} \f(t)^3 \right)   \epsilon_{i j k} \ \epsilon^{a b c} \
 A_{i +}^a A_{j +}^b A_{k +}^c  \right) \, .
\label{k4simp}
\end{equation}
The topological charge density is simply the temporal derivative of this
expression. Note that $\epsilon_{i j k} \ \epsilon^{a b c} \ A_{i +}^a A_{j
+}^b A_{k +}^c$ is proportional to the winding number density of the gauge
transformation. This means that the spatial dependence of the topological
charge density is given by the winding number density and the temporal
dependence is given by $\left( 1 -  \f (t) \right) \ \f (t) \  \f ' (t)$.
With the $g$ defined in Eq.~(\ref{hgtrans}), the $\alpha(r)$ defined in
Eq.~(\ref{alpha}) and the $\f(t)$ defined in Eq.~(\ref{fk_ramp}) the
topological charge density evaluates to
\begin{equation}
	q (r,t) = \frac{3}{d \pi r^2} \cos^2 \left( \frac{\pi (r-R)}{d} \right)
 \left( \frac{1}{4 \Delta t} - \frac{t^2}{\Delta t^3} \right) \cdot
\begin{cases}
 1 & \mbox{for} \  1 \leq t \leq  1 + \Delta t \ \  \mbox{and} \ \
   R - \frac{d}{2} \leq r \leq R + \frac{d}{2} \\
 0 & \mbox{else}
\end{cases} .
\label{topchargeres}
\end{equation}
As can easily be checked, integrating this expression ($r^2 dr$ and $dt$)
yields $Q=1$.



\section{The generalized spherical vortex on the lattice}
\label{sec:gen_latt}

We now put the generalized continuum spherical vortex of
Eq.~(\ref{vvtransgeneral}) onto the lattice with periodic boundary
conditions. Clearly, the field as given in Eq.~(\ref{vvtransgeneral}) does not
fulfill periodic boundary conditions in the temporal direction.
$\mathcal{A}_{\mu}$ vanishes for $t \to - \infty$ (and also for $r \to
\infty$) but not for $t \to \infty$. If we want to get a vanishing gauge
field $\mathcal{A}_{\mu}$ for $t \to \infty$ by gauge transforming the
field of Eq.~(\ref{vvtransgeneral}), we have to find a gauge transformation
that equals $\mathbbm{1}$ for $t \to -\infty$ and $g^\dagger$ for $t \to
\infty$. As can be shown easily by continuity arguments, there is no
continuous gauge transformation that fulfills that criteria for the $g$
given in Eq.~(\ref{hgtrans}). However, there is still the possibility to
put the field onto a lattice of infinite size. Then, one can transform the
links for  $t  \to \infty$ to unity and subsequently close the lattice by
periodic boundary conditions. Performing this procedure for the continuum
field of Eq.~(\ref{vvtransgeneral}) with $\f(t)$ given in Eq.~(\ref{fk_ramp}) yields
\begin{align}
\begin{split}
	U_{i} (x) &= \begin{cases}
 \left( g\left( \vec{r} + \hat{i} \right) \ g\left(\vec{r}\right)^{\dagger} \right)^{(t-1)/\Delta t}
 & \mbox{for} \quad 1 < t < 1 + \Delta t \\
 g\left(\vec{r} + \hat{i}\right) \ g\left(\vec{r}\right)^{\dagger}
 & \mbox{for} \quad 1 + \Delta t \leq t \leq t_g \\
 \mathbbm{1} & \mbox{else}
\end{cases} , \\
	U_{4} (x) &= \begin{cases}
 g(\vec{r})^{\dagger}  & \mbox{for} \quad t = t_g \\
 \mathbbm{1} &\mbox{else}
\end{cases} .
\label{reallattobj}
\end{split} 
\end{align}
The functions $g(\vec{r})$ and $\alpha(r)$ are again given by
Eqs.~(\ref{hgtrans}) and (\ref{alpha}).

\begin{figure}
  \centering
 		a)\includegraphics[width=0.46\columnwidth]{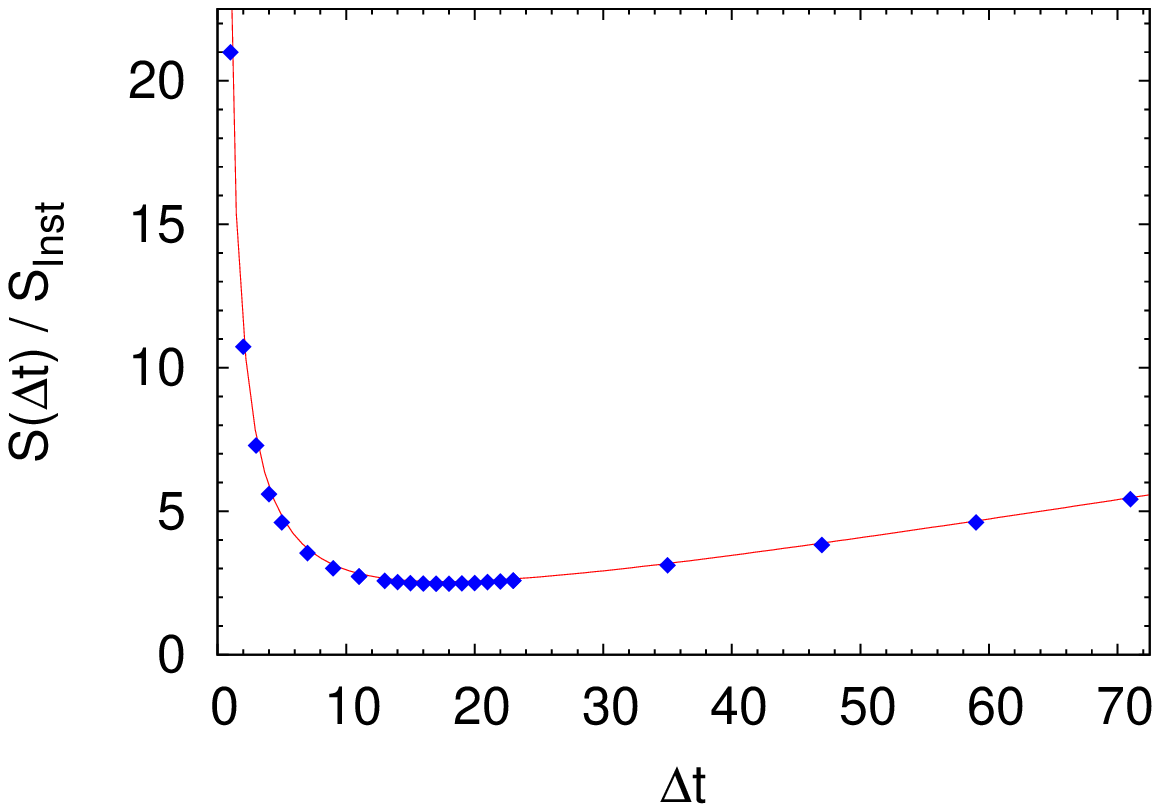}
 		b)\includegraphics[width=0.46\columnwidth]{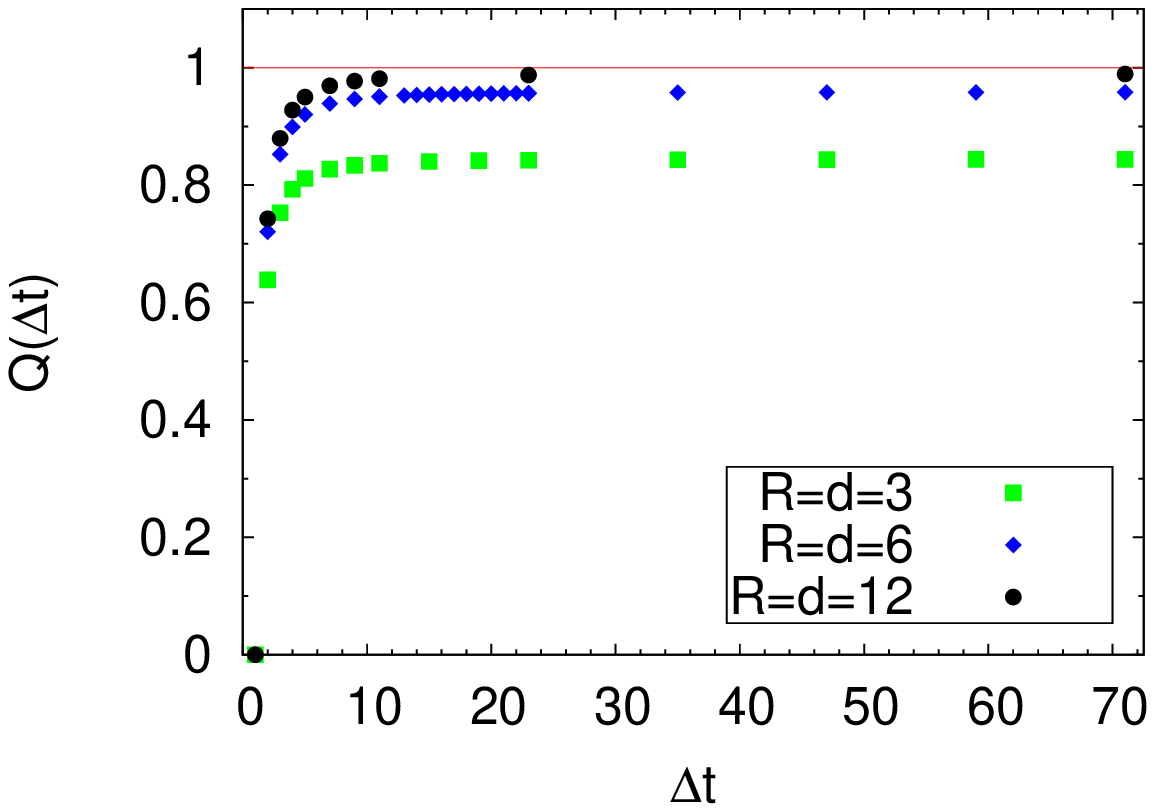}
	\caption{
a) Comparison of the gauge action (in units of the instanton action $S_{Inst}$) on the lattice, (blue) diamonds, with the gauge action of the continuum object, (red) line, given by Eq.~(\ref{eval_action_ramp_fk}) as function of the temporal extent $\Delta t$ of the spherical vortex. The radius as well as the thickness of the vortex are given by $R = d = 6$. The midpoint of the vortex is half integer.
b) Comparison of the gluonic topological charge $Q(\Delta t)$ for three
values of $R=d$. The lattice results (calculated with the plaquette
definition) are shown as (green) squares, (blue) diamonds and (black)
circles.  In the continuum, $Q(\Delta t) = 1$ for all $\Delta t$ as shown
by the (red) horizontal line. The lattice sizes $N_s^3\times N_t$ are chosen to fit the spacial and temporal extents of the spherical vortex.}
	\label{fig:s_latt_vs_ana}
\end{figure}

Let us now have a look at the gauge action and topological charge of this
generalized spherical vortex on the lattice. The gauge action as function
of the temporal extent $\Delta t$ is plotted in Fig.~\ref{fig:s_latt_vs_ana}a). Note, that the action of the lattice object matches the action of the underlying continuum object pretty well. However, it systematically underestimates the continuum value.
In Fig.~\ref{fig:s_latt_vs_ana}b) we show the topological charge density as
function of $\Delta t$ for three values of $R=d$. As one can see, the
discrepancy between the topological charge $Q$ on the lattice and in the
continuum (which is always 1) is dramatic for small temporal extent $\Delta
t$ of the spherical vortex. For $\Delta t = 1$ the topological charge of
the continuum object is not recognized. However, one still gets a zeromode
for the corresponding lattice Dirac operator, {\it i.e.}, the fermions
still see the topological charge of the underlying continuum object. This
is not unexpected as the vacuum to vacuum transition is still present in
the lattice object. To be more precise, the lattice samples the field
before and after the transition. The transition itself falls between two
time-slices. As discussed in section \ref{sec:topdencont}, the transition
carries the topological charge. Therefore, by missing the transition, one
also misses the topological charge. From Fig.~\ref{fig:s_latt_vs_ana} we
see, that also for very big $\Delta t$ (slow vacuum to vacuum transitions),
the lattice topological charge doesn't quite approach one. However,
increasing $R$, keeping $R=d$ in the example shown, increases $Q$ towards
one. Since $R$ is the only spacial scale, $1/R$ acts as the spacial lattice spacing, showing that the deviations from $Q=1$ at large $\Delta t$ ($\Delta t \gtwid 10$) are discretization effects in the spatial directions.

\begin{figure}
	\centering
		\includegraphics[width=0.5\columnwidth]{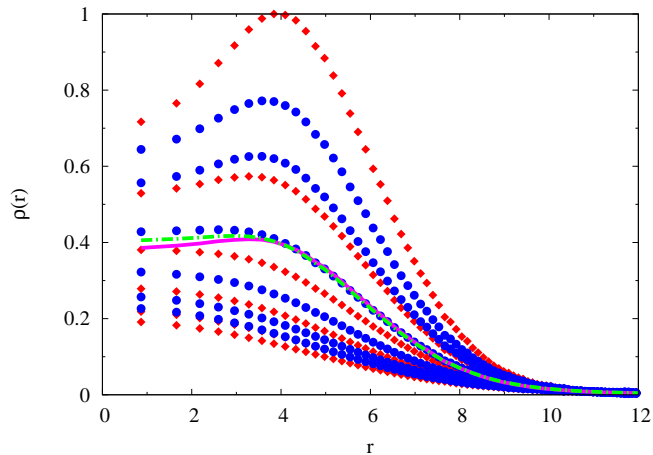}
	\caption{$\rho(r,t)$ as defined in Eq.~(\ref{def_rho_r}) with
normalization $\rho_\mathrm{max}=1$. The configuration investigated is
given by Eq.~(\ref{reallattobj}) with $\Delta t = 1$, (red) diamonds, and
$\Delta t = 5$, (blue) circles. The parameters of $\alpha(r)$ are given by
$R=d=6$ in both cases. 
The calculations have been performed on a $24^3 \times 12$
lattice. The different curves belong to different values of $t$. For
$\Delta t = 1$ we have $t \in \{1\eqdef2,3\eqdef12,\ldots,7\eqdef8\}$ and for $\Delta t = 5$ we have $t \in \{3\eqdef4,5\eqdef2,\ldots,9\eqdef10\}$. The highest curve
corresponds to the first pair of $t$-values, the lowest curve to the last pair
of $t$-values. In between the curves 
fall monotonically. The solid (magenta) line represents the
average over $t$ for $\Delta t = 1$, the dash-dotted (green) line
represents the same for $\Delta t = 5$.}
	\label{fig:htrans2_6_rad_dep}
\end{figure}

To conclude this section we briefly discuss how the fermionic results
depend on the temporal extent of the vortex. For the fermion fields we use
antiperiodic boundary conditions in the temporal direction and periodic
boundary conditions in the spatial directions. With these boundary
conditions, the Dirac operator possesses always exactly one zeromode for
gauge configurations of the form of Eq.~(\ref{reallattobj}), independent of
the values of the parameters. This zeromode is more or less located at the
vortex. For the calculations, we used the overlap Dirac operator. With
the described setting, we see that, as expected, the scalar density of the zeromode is smeared out in the temporal direction for the vortices with bigger
temporal extent. The spatial localization, however, is more or less
independent of the temporal extent of the vortex. Thus, all what really
matters is that the vacuum to vacuum transition occurs, not how fast it
occurs.
Let us discuss this in a little bit more detail. First, we note that the
scalar density of the zeromode is distributed, almost perfectly,
spherically symmetric around the midpoint $\vec{r}_0$ of the vortex.
Keeping this in mind, the discussion of the localization in 3 dimensional
space reduces to a discussion of the localization in $|\vec{r} -
\vec{r}_0|$. Thus we study
\begin{equation}
	\rho(r,t) = \frac{1}{N(r)} \sum_{\vec{r}}{ \delta(r, |\vec{r} - \vec{r}_0|) \
 \psi_0^{\dagger} (\vec{r},t) \psi_0 (\vec{r},t) } \quad \mbox{with} \quad
N(r) = \sum_{\vec{r}}{ \delta(r, |\vec{r} - \vec{r}_0|) } \, .
\label{def_rho_r}
\end{equation}
In other words, the quantity $\rho(r,t)$ stands for the mean value of the
scalar density of the zeromode, calculated for points belonging to the same
$|\vec r-\vec r_0|$ and $t$. In Fig.~\ref{fig:htrans2_6_rad_dep} the results for $\rho(r,t)$ for generalized spherical vortices with $\Delta t = 1$ and
$\Delta t = 5$ are compared. One can see the similarity between the result
for the two different $\Delta t$.

\section{Discussion and Outlook}

In this paper, we identified the continuum object corresponding to the
previously considered spherical vortex as a vacuum to vacuum transition in
temporal direction. The discrepancy between the gluonic lattice topological
charge and the index of the lattice Dirac operator, described in previous
papers, turned out to be a discretization effect in the temporal direction.
Starting from the continuum spherical vortex, we constructed a generalized
spherical vortex on the lattice.  We demonstrated the similarity to the
original spherical vortex by using fermions as probes. We also showed that,
with an appropriate choice of parameters, the action of the generalized
spherical vortex can be quite small, as small as about 5/3 of the
one-instanton action. For more details see~\cite{Schweigler:2012}.

It is known that topological charge contributions from center vortices, due
to intersection (and writhing) points, emerge from vortex structures lying
in a single U(1) subgroup~\cite{Reinhardt:2002cm}. Color rotations in such
structures are suppressed by the action. For example, intersections of
vortices with orthogonal color structures give maximally negative
plaquettes and are thus suppressed in the continuum limit. However, this
does not mean that the whole vortex structure can be gauge transformed to a
single U(1) subgroup.  We conjecture, for entropic reasons, that in the
confined phase the color structure of center vortices contributes to the
topological charge and chiral symmetry breaking. 

Further investigations should therefore aim at quantifying these kinds of
contributions in Monte Carlo generated configurations. First measurements show
that the total vortex surface definitely covers the full $S^3$, {\it i.e.}
vortex plaquettes are distributed uniformly among the entire $SU(2)$ color
palette. However, color vectors are not gauge invariant and therefore the number
of full coverings is not so easy to measure, even though it is a gauge
invariant, topological quantity. Further, writhing points 
dominate the topological susceptibility~\cite{Bertle:2001xd,Engelhardt:2000wc},
even in $SU(3)$~\cite{Engelhardt:2010ft}, and therefore color contributions
might only play a sub-dominant role like intersection points. But the colorful spherical vortex may act as an ansatz for a model of chiral symmetry breaking, as it shows properties similar to instantons~\cite{Hollwieser:2013}.

\bibliography{literatur}

\end{document}